\documentclass[conference]{IEEEtran}
\usepackage{cite,graphicx,amssymb,amsmath,color,textcomp}
\usepackage{extarrows,multirow,multicol}
\usepackage{bm}
\usepackage{subeqnarray}
\usepackage{float}
\usepackage[noend]{algorithmic}
\usepackage{algorithm}
\usepackage{comment}
\usepackage{amssymb}
\usepackage{subfigure}
\usepackage{soul}

\usepackage{enumitem}

\def \hd {\mathcal{H}_1}
\def \hn {\mathcal{H}_0}

\newlength{\figwidth}
\IEEEoverridecommandlockouts

\begin{document}
\bstctlcite{IEEEexample:BSTcontrol}
\title{
Uniform Planar Array Based Weighted Cooperative Spectrum Sensing for Cognitive Radio Networks
}
\author{
\IEEEauthorblockN{Charith~Dissanayake\IEEEauthorrefmark{1}, Saman~Atapattu\IEEEauthorrefmark{1},
Prathapasinghe~Dharmawansa\IEEEauthorrefmark{2}, 
Jing ~Fu\IEEEauthorrefmark{1}, \\
Sumei Sun\IEEEauthorrefmark{3}, and 
Kandeepan~Sithamparanathan\IEEEauthorrefmark{1}}
 \IEEEauthorblockA{
\IEEEauthorrefmark{1}Department of Electrical and Electronic Engineering, RMIT University, Victoria, Australia\\
\IEEEauthorrefmark{2}Centre for Wireless Communications,
University of Oulu, Finland\\
\IEEEauthorrefmark{3}Institute for Infocomm Research, Agency for Science, Technology and Research, Singapore
\IEEEauthorblockA{Email:\IEEEauthorrefmark{1}\{charith.dissanayake,\,saman.atapattu,\,jing.fu,\,kandeepan.sithamparanathan\}@rmit.edu.au;\,\\
\IEEEauthorrefmark{2}pkaluwad24@univ.yo.oulu.fi;\,
\IEEEauthorrefmark{3}sunsm@i2r.a-star.edu.sg}
}
\vspace{-1cm}

 

}
\maketitle
\begin{abstract} 
Cooperative spectrum sensing (CSS) is essential for improving the spectrum efficiency and reliability of cognitive radio applications. Next-generation wireless communication networks increasingly employ uniform planar arrays (UPA) due to their ability to steer beamformers towards desired directions, mitigating interference and eavesdropping. However, the application of UPA-based CSS in cognitive radio remains largely unexplored. This paper proposes a multi-beam UPA-based weighted CSS (WCSS) framework to enhance detection reliability, applicable to various cognitive radio networks, including cellular, vehicular, and satellite communications.
We first propose a weighting factor for commonly used energy detection (ED) and eigenvalue detection (EVD) techniques, based on the spatial variation of signal strengths resulting from UPA antenna beamforming.
We then analytically characterize the performance of both weighted ED and weighted EVD by deriving closed-form expressions for false alarm and detection probabilities. Our numerical results, considering both static and dynamic user behaviors, demonstrate the superiority of WCSS in enhancing sensing performance compared to uniformly weighted detectors. 
\end{abstract}
\section{Introduction} \label{S1}
The evolution of wireless technologies toward 6G networks offers novel opportunities for intelligent, high-speed communications with high-reliability and low latency. However, spectrum scarcity remains a significant challenge~\cite{tataria20216g}. Cognitive radio (CR) technology has emerged as a promising solution to address spectrum scarcity~\cite{Haykingfirst}. The effectiveness of CR depends heavily on spectrum sensing (SS) techniques that dynamically identify and utilize white spaces for communication. Traditional SS approaches face challenges due to propagation phenomena such as noise, multi-path fading, and shadowing effects~\cite{akyildiz2006next,van_trees_detection_2001}. Cooperative spectrum sensing (CSS) strategies have proven effective in overcoming these limitations by fusing spatially distributed signal observations~\cite{atapattu2011energy,Guo2018tcom}.

Simultaneously, various advanced forms of multiple-input multiple-output (MIMO) technologies have been introduced to wireless communications, enhancing user capabilities~\cite{zhang2020mimo,GLRTMOMOCSI}. MIMO technologies such as massive MIMO~\cite{dey2020wideband,mMIMOtest}, reconfigurable intelligent surfaces (RIS)~\cite{ge2024ris}, Uniform Planar Arrays (UPA)~\cite{balanisbook,mailloux2017phased}, Uniform Linear Arrays (ULA) which is a subset of UPA~\cite{balanisbook}, holographic MIMO (HMIMO), and stacked intelligent metasurfaces (SIM)\cite{an2023stacked} offer higher throughput, multi-user support, and beam-steering capabilities. In \cite{zhang2020mimo,GLRTMOMOCSI,dey2020wideband,mMIMOtest,ge2024ris,MIMOcsi}, MIMO, massive MIMO, RIS applied to SS in CR improve not only communication capabilities but also the effectiveness of CSS in CR for better spectrum utilization.


Specifically, an SS scheme using a Gaussian mixture noise model and kernelized test statistic for MIMO was proposed in \cite{zhang2020mimo}, while \cite{GLRTMOMOCSI} presented an SS approach for MIMO  using a generalized LRT-based test statistic with CSI uncertainty. Similarly, a RIS-assisted CSS was proposed in \cite{ge2024ris}, optimizing the phase shift matrix to maximize detection probability. 
Furthermore, massive MIMO-based decision fusion is explored in \cite{dey2020wideband}, with large number of antennas at fusion center (FC). This approach incorporates wideband CSS with Orthogonal Frequency Division Multiplexing for data reporting to the FC and results of similar work is carried out on a testbed in~\cite{mMIMOtest}. Likewise, MIMO-based communication from sensors to FC explored in \cite{MIMOcsi}. Additionally, a relay-based CSS introduced in~\cite{atapattu2009relay}, where relay nodes convey the primary user signal to an FC. The SS methods discussed above employ various detection techniques, including energy detection (ED), eigenvalue detection (EVD), and likelihood detection~\cite{atapattu2011energy,atapattu2014energy,Chamain2020it,mcwhorter2023passive}.

While MIMO, massive MIMO, RIS, and virtual MIMO relay networks have been studied for SS in CR, there is a notable research gap in exploring other MIMO-like technologies such as HMIMO, SIM, and UPA. UPAs recently gained significant attention for their structural simplicity, spatial efficiency, three-dimensional beamforming capability~\cite{sultan2020fast,balanisbook,mailloux2017phased}. Moreover, their modified architecture can represent HMIMO, RIS, and SIM, despite each having unique features. These capabilities make UPAs particularly suitable for various applications, including terrestrial and non-terrestrial communications like satellite communications, cellular networks, and vehicular ad-hoc networks (VANETs). {\it This work addresses the research gap by considering UPA-based CSS for CR applications, providing novel contributions for future research}.

The key contributions of this research are: 1) By leveraging the spatial variation of signal strength across users due to UPA multiple beamforming, we propose a weighted CSS scheme for energy and eigenvalue detections; 2) We derive rigorous expressions for data fusion-based WCSS for false-alarm and detection probabilities using probability theory and random matrix theory; and 3) We validate the framework through comprehensive numerical analysis, demonstrating enhanced sensing capabilities and its applicability to UPA-based general (static) and vehicular (dynamic) wireless networks. Moreover, the proposed scheme and analytical framework can be readily utilized with minor modifications for HMIMO and SIM-based CR applications, which have not been considered so far. 
\begin{figure}[t!]
  \centering
  \includegraphics[width=0.35\textwidth]{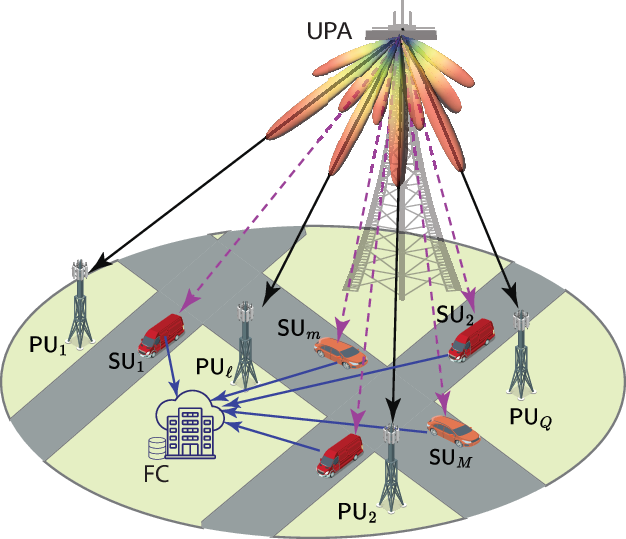}
  \caption{System model of a VANET with \textsf{PUs}, \textsf{SUs}, and a UPA for downlink.}\label{system_model}
\end{figure}
\begin{figure}[t!]
  \centering
  \includegraphics[width=0.35\textwidth]{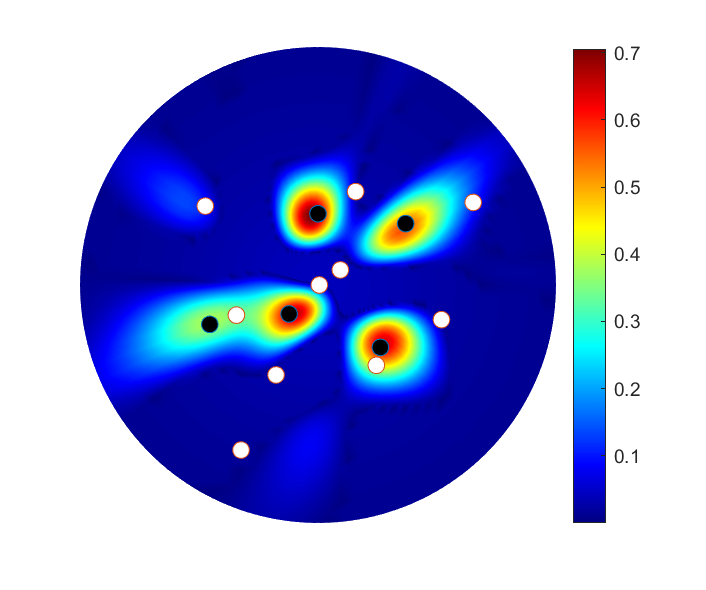}
  \caption{Spatial variation of SNR due to UPA beamforming, as a heatmap.}\label{heatmap}
\end{figure}
\section{System Model}
\subsection{Network Model}
We consider a communication network comprising \( Q \) primary users (\textsf{PU}s) denoted as \(\textsf{PU}_\ell\) for \(\ell \in \{1, \dots, Q\}\), and \( M \) secondary users (\textsf{SU}s) represented by \(\textsf{SU}_m\) for \( m \in \{1, \dots, M\}\). The network employs a UPA antenna for downlink communication with the \textsf{PU}s. The UPA consists of \( L \) elements arranged as \( L_x \) elements along the \( x \)-axis with spacing \(\Delta x\) and \( L_y \) elements along the \( y \)-axis with spacing \(\Delta y\). Each \textsf{SU} is equipped with a single omnidirectional antenna, while the UPA generates multiple beams directed at the \textsf{PU}s. The \textsf{SU}s forward their sensing information to an FC for a centralized decision on the presence of communication with the \textsf{PU}s. This network setup is representative of a VANET, as shown in Fig.~\ref{system_model}. The UPA and \textsf{PU}s are assumed to have knowledge of each other’s locations.

\subsection{Signal Model}
The power radiation of a UPA is determined by its array factor (AF). For a UPA with its primary beam directed toward \textsf{PU}s at elevation angles \(\bm{\theta} = [\theta_1, \dots, \theta_Q]\) and azimuth angles \(\bm{\phi} = [\phi_1, \dots, \phi_Q]\), the array factor \( AF \) toward each \(\textsf{SU}_m\) at elevation and azimuth angles (\(\theta_m,\,\phi_m\)) is given as \cite{balanisbook,mailloux2017phased}
\begin{align}
    &AF(\theta_m, \phi_m,\bm{\theta},\bm{\phi},\mathbf{a}) =4\sum_{\ell=1}^{Q}a_{\ell} \sum_{i=1}^{L_x/2}  \sum_{j=1}^{L_y/2}e_{i,j}\nonumber\\
    &\times\cos\left[(2i - 1)\beta_1(\theta_m, \phi_m)\right]
    \cos\left[(2j - 1)\beta_2(\theta_m, \phi_m)\right]
\end{align}
where $\beta_1(\theta_m,\phi_m) = (\sin \theta_m \cos \phi_m - \sin \theta_{\ell} \cos \phi_{\ell})\pi \Delta x f_c/c $ and $\beta_2(\theta_m, \phi_m) =(\sin \theta_m \sin \phi_m - \sin \theta_{\ell}\sin \phi_{\ell})\pi \Delta y f_c/c $. 
Here, \( e_{i,j} \) denotes the amplitude excitation of the antenna element located at \((i, j)\) in the array. The vector \( \mathbf{a} = [a_1, \dots, a_Q] \in \mathbb{R}^{1 \times Q} \) represents peak amplitude ratios, where \( a_\ell \) is the ratio for the \( \ell \)-th main beam. The communication operates at a frequency \( f_c \), with \( c \) as the speed of light. Given that \(\bm{\theta}, \bm{\phi}, \mathbf{a}\) are fixed in the system model, we denote the array factor toward \((\theta_m, \phi_m)\) is as \( AF(\theta_m, \phi_m) \). The radiated electric field in the direction \((\theta_m, \phi_m)\) is then given by \( E(\theta_m, \phi_m) \cdot AF(\theta_m, \phi_m) \), where \( E(\theta_m, \phi_m) \) denotes the element factor for that direction. 
The communication between the UPA and \textsf{PU}s is either active or idle during the detection period, represented by the hypotheses \(\mathcal{H}_1\) (presence of \textsf{PU}s) and \(\mathcal{H}_0\) (absence of \textsf{PUs}).


The unknown transmission signal vector is \(\mathbf{s} = [s[1], \dots, s[K]] \in \mathbb{C}^{1 \times K}\), comprising \( K \) samples with \(\mathbf{s} \sim \mathcal{CN}(\mathbf{0}_K, \sigma_s^2 \mathbf{I}_K)\) and transmission power \( p \). The noise vector \(\mathbf{n}_m = [n_m[1], \dots, n_m[K]] \in \mathbb{C}^{1 \times K}\) contains i.i.d. samples \(n_m[k] \sim \mathcal{CN}(0, \sigma_n^2)\).  Under this, the received signal vector at \(\textsf{SU}_m\), \(\mathbf{y}_m = [y_m[1], \dots, y_m[K]] \in \mathbb{C}^{1 \times K}\), is given by 
\begin{align}\label{original_hypothesis}
\mathbf{y}_m =  
\begin{cases}
     \mathbf{n}_m,  & \mathcal{H}_0, \\
     \alpha_m\mathbf{s} + \mathbf{n}_m, & \mathcal{H}_1,
\end{cases}
\end{align}
where $\alpha_m \hspace{-0.5mm}=\hspace{-0.5mm} \sqrt{\hspace{-0.5mm} (pGc^2\hspace{-0.5mm} |\hspace{-0.5mm}AF(\hspace{-0.5mm}\theta_m, \phi_m\hspace{-0.5mm})E(\hspace{-0.5mm}\theta_m, \phi_m\hspace{-0.5mm})\hspace{-0.5mm} |^2\hspace{-0.5mm} /\hspace{-0.5mm}(\hspace{-0.5mm}4\pi f_c r_m)^2\hspace{-0.5mm})}h_m$
with \( G \) representing the combined antenna gain, \( h_m \in \mathbb{C} \) denoting the channel gain, and \( r_m \) representing the distance between the UPA and \(\textsf{SU}_m\). 
The received signal vector follows a complex Gaussian distribution under both hypotheses, expressed as
\begin{align}\label{y_distribtuion}
\mathbf{y}_m \sim 
\begin{cases}
    \mathcal{CN}(\mathbf{0}, \sigma_n^2 \mathbf{I}_K) & \quad \mathcal{H}_0 \\
    \mathcal{CN}(\mathbf{0}, (|\alpha_m|^2\sigma_s^2 + \sigma_n^2) \mathbf{I}_K). & \quad \mathcal{H}_1
\end{cases}
\end{align}

\section{Weighted Cooperative Spectrum Sensing}\label{WCSS}
Cooperative spectrum sensing improves decision accuracy by leveraging multiple \textsf{SU}s to sense the spectrum. Raw signal samples are transmitted to the FC for combined analysis of primary communication. Due to the UPA antenna's distinct radiation pattern with beams directed at \textsf{PU}s, the received SNR varies spatially, as shown in Fig.~\ref{heatmap}. 
To enhance CSS decision accuracy, it is crucial to prioritize \(\textsf{SU}s\) with higher SNR by assigning appropriate weights to their samples. We propose a weighted CSS scheme that applies a normalized weight \( w_m \geq 0 \) to each sample, with the weighting vector \( \mathbf{w} = [w_1, \dots, w_M] \) satisfying \( \sum_{m=1}^{M} w_m = 1 \).
Various methods exist to determine optimal or sub-optimal weighting factors in CSS (e.g., Neyman-Pearson\cite{van_trees_detection_2001}). We choose a weighting factor representing all the user-specific parameters that emphasizes reliable decisions as
\begin{align}
    w_m = \frac{|\alpha_m|^2}{\sum_{i=1}^{M}|\alpha_i|^2},\quad m = 1, \dots, M.
\end{align}
In what follows, we use \(w_m\) for both weighted ED (WED) and weighted eigenvalue detection (WEVD).
\subsection{Weighted Energy Detector (WED)}  
The received signal samples at the FC are weighted by the square root of the corresponding weights to obtain the weighted average energy. The received weighted signal \(\mathbf{Z} \in \mathbb{C}^{M \times K}\), for \(M\) users, each with \(K\) samples, is given by
\begin{align}\label{signal_matrix}
    \mathbf{Z} = \mathbf{W}^\frac{1}{2}\mathbf{Y}
\end{align}
where \( \mathbf{Y} = \begin{bmatrix} 
\mathbf{y}_1^T, \dots, \mathbf{y}_M^T
\end{bmatrix}^{T} \in \mathbb{C}^{M \times K} \) and \( \mathbf{W} = \text{diag}(w_1, w_2, \dots, w_M) \in \mathbb{R}^{M \times M} \) are the received signal and the diagonal matrix of the weighting factors, respectively. The total average energy of \(K\) samples from \(M\) users, \(\Lambda_{\text{WED}}\), is the test statistic given by
\begin{align}
\Lambda_{\text{WED}} = \frac{1}{K}\sum_{m=1}^M w_m |\mathbf{y}_m|^2 \underset{\hn}{\overset{\hd}{\gtrless}}\tau. 
\end{align}


\subsection{Weighted Eigenvalue Detector (WEVD)}
The received signal sample matrix at the FC, as given in \eqref{signal_matrix}, is used to calculate the unnormalized sample covariance matrix \(\mathbf{R} = \mathbf{Z} \mathbf{Z}^H\). The test statistic \(\Lambda_{\text{WEVD}}\) is then the maximum eigenvalue of \(\mathbf{R}/K\), expressed as
\begin{align}\label{eq_evd_cov}
\Lambda_{\text{WEVD}} = \lambda_{\max}\left( \mathbf{Z} \mathbf{Z}^H/K\right)  \underset{\hn}{\overset{\hd}{\gtrless}}\tau.
\end{align}
We now proceed to evaluate the performance of proposed WCSS techniques. 

 \begin{figure*}[ht]
    \centering
    \subfigure[Comparison of WCSS and uniform weighted\label{ROC_data_fusion}]
    {\includegraphics[width=0.32\textwidth]{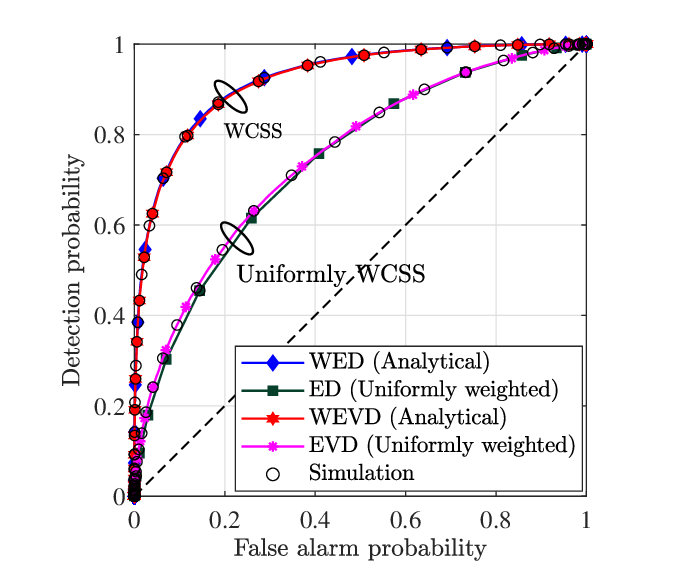}} 
    \subfigure[Impact of \textsf{SU} count $M=5,10$\label{ROCdata_users}]{\includegraphics[width=0.32\textwidth]{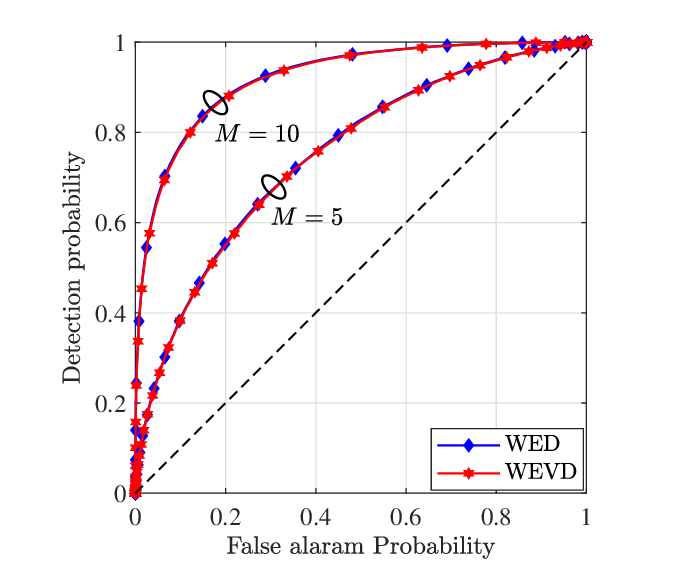}} 
    \subfigure[Mobility with different weight update intervals \label{ROCdata_mobility}]{\includegraphics[width=0.32\textwidth]{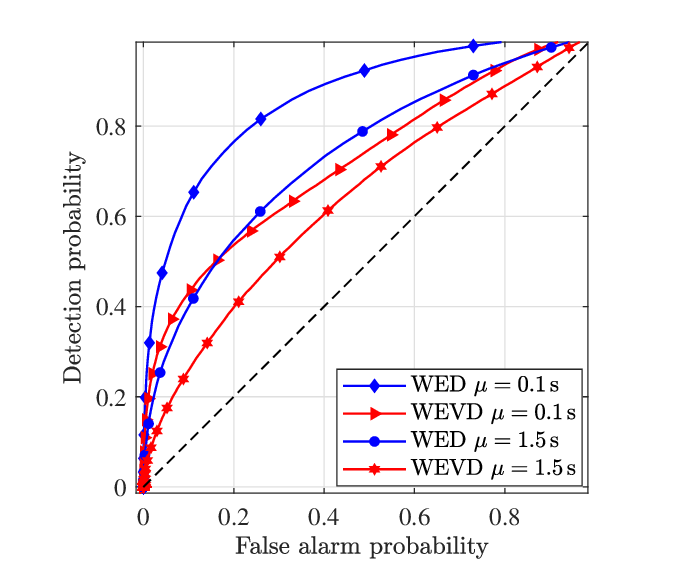}}
    \caption{ROC analysis under static and mobility scenarios with \( L = 64\).}
\end{figure*}
\begin{figure*}[ht]
    \centering
    \subfigure[Comparison for \( L = 64 \) and \( L = 128 \) elements\label{ROCdatadecisionvant}]{\includegraphics[width=0.32\textwidth]{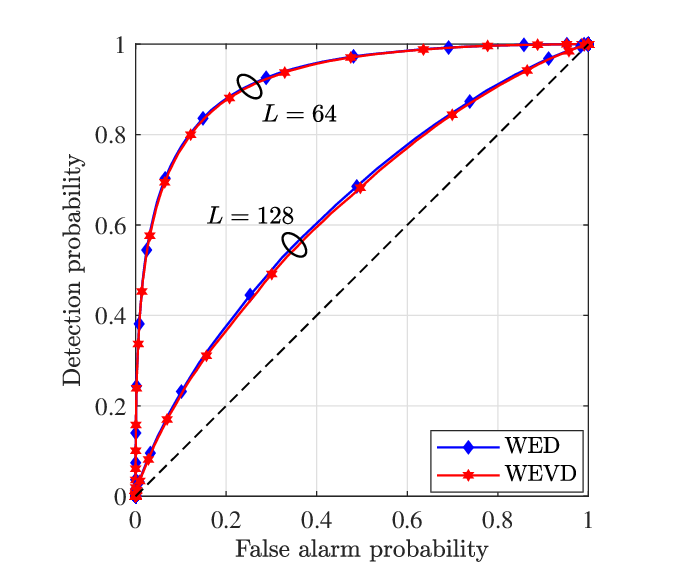}}
    \subfigure[Heatmap of received signal strength \( L = 64 \)\label{Heatmap16}]
    {\includegraphics[width=0.32\textwidth]{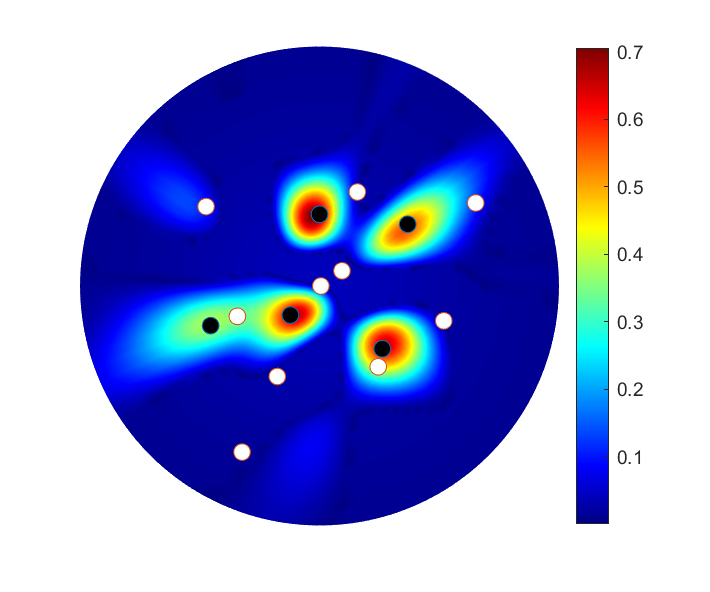}} 
    \subfigure[Heatmap of received signal strength \( L = 128 \)\label{Heatmap128}]{\includegraphics[width=0.32\textwidth]{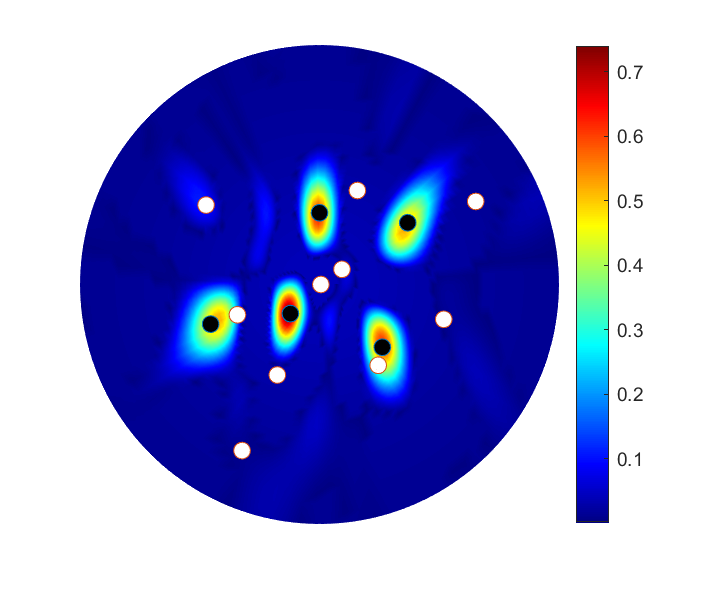}} 
    \caption{Detection performance comparison for different antenna configurations \( L = 64 \) and \( L = 128 \) with \( M = 10 \).}
\end{figure*}
\section{Detection Performance of WCSS} \label{performance}
Here we characterize the performance of WCSS systems with respect to the detectors introduced. The performance of a detector is evaluated by probability of false alarm $ P_f(\tau)~=~\Pr(\Lambda>\tau|\hn)$ and probability of detection \(P_d(\tau)~=~\Pr(\Lambda>\tau|\hd)\). Therefore, we derive the expressions for $P_f(\tau)$ and $P_d(\tau)$ here.

\subsection{Performance of WED}
The test statistic of signal samples at FC is given by
     $\Lambda_{\text{WED}}  = \frac{1}{K}\sum_{m=1}^{M}w_m\sum_{k=1}^{K}|y_m[k]|^2$,  where
 \(w_m \sum_{k=1}^{K} |y_m[k]|^2\) represents the weighted energy of signal samples from \(\textsf{SU}_m\). To compute \(\Lambda_{\text{WED}}\), we first evaluate the distribution of \(w_m \sum_{k=1}^{K} |y_m[k]|^2\). To facilitate further analysis, the PDF of a gamma distributed RV with shape \(k\) and scale \(\theta\) is given by  $f(x)= x^{k - 1} e^{-x/\theta}/(\Gamma(k) \theta^k),\, x\geq 0$. 

Since \( y_m[k] \) follows a complex Gaussian distribution under \( \mathcal{H}_0 \) and \( \mathcal{H}_1 \), \( |y_m[k]|^2 \) follows an exponential distribution with rate parameter \( \sigma_n^2 \) under \( \mathcal{H}_0 \) and \( \sigma_n^2(\gamma_m + 1) \) under \( \mathcal{H}_1 \), where \( \gamma_m = \frac{|\alpha_m|^2 \sigma_s^2}{\sigma_n^2} \) is the SNR at \( \textsf{SU}_m \). Therefore, \( w_m \sum_{k=1}^{K} |y_m[k]|^2 \) follows a Gamma distribution with shape \( K \) and scale \( \sigma_n^2 w_m \) under \( \mathcal{H}_0 \) and \( \sigma_n^2 w_m (\gamma_m + 1) \) under \( \mathcal{H}_1 \).
Here, with non-identical parameters for each \( m \), the PDF of \( \Lambda_{\text{WED}} \) under \( \mathcal{H}_0 \) is the sum of independent, non-identical Gamma RVs, which is given by \cite{sum_of_gamma_al}
\begin{align}
     \hspace{-2mm}f_{\Lambda_{\text{WED}}}(\hspace{-0.5mm} x\hspace{-0.5mm})\hspace{-0.5mm} = \hspace{-0.5mm}K\hspace{-1.5mm}\prod_{m=1}^{M} \hspace{-2mm}\left(\hspace{-1mm}\frac{w_1}{w_m}\hspace{-1mm}\right)^{\hspace{-1mm}K} \hspace{-1mm}\sum_{j=0}^{\infty} \hspace{-1mm}\frac{\delta_j (K x )^{KM + j - 1} e^{\frac{-K x}{\sigma_n^2w_1}}}{(\sigma_n^2w_1)^{(KM + j)} \Gamma \hspace{-0.5mm}\left(KM\hspace{-1mm}+\hspace{-1mm}j\right)}\label{pdf_sumof_gamma_new}
\end{align}
where \( w_1 = \min\{w_1, \dots, w_M\} \), \( \delta_0 = 1 \), and
\begin{align}\label{delta}
    \delta_{j+1}=\frac{K}{j+1} \sum_{i=1}^{j+1}\hspace{-1mm} \left[ \sum_{q=1}^{M} \left( \hspace{-1mm}1\hspace{-1mm} -\hspace{-1mm} \frac{w_1}{w_{q}}\hspace{-1mm} \right)^i \right]\hspace{-1mm} \delta_{j+1-i}, j = 0, 1,\cdots
\end{align}
By using \eqref{pdf_sumof_gamma_new}, we can calculate $P_{f,\text{WED}}(\tau)$ as
\begin{align}\label{pfwed}
    \hspace{-1mm}P_{f,\text{WED}}(\hspace{-0.5mm}\tau\hspace{-0.5mm})\hspace{-1mm} =&1\hspace{-1mm}-\hspace{-1mm}\prod_{m=1}^{M}\hspace{-0.7mm} \left(\hspace{-1mm}\frac{w_1}{w_m}\hspace{-1mm}\right)^{\hspace{-1mm}K}\hspace{-1mm} \sum_{j=0}^{\infty}\hspace{-0.2mm} \delta_j\hspace{-0.2mm}\frac{\gamma(KM + j,\frac{K\tau}{\sigma_n^2w_1})}{\Gamma\left(KM + j\right)} 
\end{align}
where \( \gamma(s, x) = \int_0^x t^{s-1} e^{-t} \, {\rm d}t \). 
Under \( \mathcal{H}_1 \), the weighted energies are neither identical nor independent across \textsf{SUs}, so \( \Lambda_{\text{WED}} \) is expressed a sum of non-independent and non-identically distributed (n.i.n.i.d.) gamma RVs, given by \cite{sum_of_gamma_al}
\begin{align}
     \hspace{-2mm}f_{\Lambda_{\text{WED}}}(\hspace{-0.1mm} x\hspace{-0.1mm})\hspace{-1mm} = \hspace{-1mm}K\hspace{-0.5mm}\prod_{m=1}^{M} \hspace{-0.5mm}\left(\hspace{-0.5mm}\frac{\lambda_1}{\lambda_m}\hspace{-1mm}\right)^{\hspace{-1mm}K} \hspace{-1mm}\sum_{j=0}^{\infty} \frac{\delta_j (\hspace{-0.5mm}K \hspace{-0.5mm}x\hspace{-0.5mm} )^{KM + j - 1} e^{\frac{-K x}{\lambda_1}}}{\lambda_1^{KM +\hspace{-0.2mm} j} \Gamma \hspace{-0.5mm}\left(KM \hspace{-1mm}+\hspace{-1mm}j\right)}\label{pdf_sumof_correlatedgamma_new}
\end{align}
where \( \lambda_1 = \min\{\lambda_1, \dots, \lambda_M\} \) be the smallest eigenvalue of the matrix \( \mathbf{A} = \mathbf{DC} \), where \( \mathbf{D} = \text{diag}(\sigma_n^2 w_1 (\gamma_1 + 1), \dots, \sigma_n^2 w_M (\gamma_M + 1)) \) and \( \mathbf{C} \) is a positive definite matrix with elements \( C_{i,j} = \sqrt{\rho_{i,j}} \) for \( i \neq j \), and \( C_{i,i} = 1 \). Here, \( \rho_{ij} \) is the correlation coefficient between the average weighted energy of the \( i \)th and \( j \)th \textsf{SU}s, which can be computed as
\begin{align*}
    \rho_{ij} = \rho_{ji} = \frac{\sigma_s^4|{\alpha_i}|^2|{\alpha_j}|^2}{(|\alpha_i|^2|\alpha_j|^2\sigma_s^4 + \sigma_s^2\sigma_n^2 (|\alpha_i|^2+|\alpha_j|^2)+ \sigma_n^4)}.
\end{align*}
Further, the coefficients \( \delta_k \) can be recursively calculated using \eqref{delta}, replacing \( w_1 \) and \( w_{\ell} \) with \( \lambda_1 \) and \( \lambda_{\ell} \), respectively. By using \eqref{pdf_sumof_correlatedgamma_new}, the detection probability \( P_{d,\text{WED}}(\tau) \) is evaluated as
\begin{align}\label{pdwed}
    \hspace{-1mm}P_{d,\text{WED}}(\hspace{-0.5mm}\tau\hspace{-0.5mm})\hspace{-1mm} =&1\hspace{-1mm}-\hspace{-1mm}\prod_{m=1}^{M}\hspace{-1.7mm} \left(\hspace{-1mm}\frac{\lambda_1}{\lambda_m}\hspace{-1mm}\right)^{\hspace{-1mm}K}\hspace{-1mm} \sum_{j=0}^{\infty}\hspace{-0.2mm} \delta_j\hspace{-0.2mm}\frac{\gamma(KM\hspace{-1mm} +\hspace{-1mm} j,\frac{K\tau}{\lambda_1})}{\Gamma\left(KM + j\right)}. 
\end{align}
\subsection{Performance of WEVD}
For performance analysis of WEVD, the received signal matrix at the FC is expressed as 
\begin{align}\label{received_matrix}
    \mathbf{Z} =
    \begin{cases}
    \mathbf{W}^{1/2}\mathbf{N}&\quad \hn\\
    \mathbf{W}^{1/2}\mathbf{G}\mathbf{1}\mathbf{s} + \mathbf{W}^{1/2}\mathbf{N}&\quad \hd
    \end{cases}
\end{align}
where \( \mathbf{G} = \text{diag}(\alpha_1, \dots, \alpha_M) \in \mathbb{R}^{M \times M} \) and \( \mathbf{N} = \begin{bmatrix} \mathbf{n}_1^T, \mathbf{n}_2^T, \dots, \mathbf{n}_M^T \end{bmatrix}^T \in \mathbb{C}^{M \times K} \) represents the noise matrix. Moreover, $\mathbf{1}^{M\times1}$ denotes a column vector of 1s. Consequently, the matrix \( \mathbf{R} \) derived from \eqref{received_matrix} follows a central correlated complex Wishart distribution, defined as \cite{james}
\begin{align} \label{weighted_CW}
\mathbf{R}
\sim 
\begin{cases}
    \mathcal{CW}_M(K,\sigma_n^2\mathbf{W}) &  \mathcal{H}_0 \\
    \mathcal{CW}_M(K,\sigma_s^2\mathbf{ W}^{\frac{1}{2}}\mathbf{G}\mathbf{11}^T\mathbf{G}^H \mathbf{W}^{\frac{1}{2}}+ \sigma_n^2\mathbf{W}) &   \mathcal{H}_1. 
\end{cases}
\end{align}
Under $\hn$, $P_{f,\text{WEVD}}(\tau)$ evaluated with the help of \cite{chiani2017probability} as
\begin{align}\label{pfevd}
    P_{f,\text{WEVD}}(\tau) \hspace{-1mm}=\hspace{-1mm} 1\hspace{-1mm}-\hspace{-1mm} \frac{\det[\Psi(K\tau)]}{\prod_{i < j}^{M} (w_i\hspace{-1mm} -\hspace{-1mm} w_j) \hspace{-1mm}\prod_{i=1}^{M}\hspace{-1mm} w_i^{K-M+1} \hspace{-0.5mm}(\hspace{-0.5mm}K\hspace{-1mm}-\hspace{-1mm}i)!}.
\end{align}
where \( \Psi(K\tau) \) is an \( M \times M \) matrix with elements
\begin{align*}
    \Psi_{i,j}(x) = w_j^{K - i + 1} \, \gamma\hspace{-0.5mm} \left( K - i + 1, x/(\sigma_n^2w_j) \right)
\end{align*} 
with \( w_1\hspace{-1mm} > w_2 > \dots > w_M \), and \( \det[\cdot] \) denotes the determinant. 
Under $\hd$, $P_{d,\text{WEVD}}(\tau)$ evaluated with the help of \cite{chiani2017probability} as
\begin{align}\label{pdevd}
    P_{d,\text{WEVD}}(\tau)\hspace{-1mm} =\hspace{-1mm} 1\hspace{-1mm}-\hspace{-1mm} \frac{\det[\Phi(K\tau)]}{ \prod_{i < j}^{M} (\psi_i - \psi_j)\hspace{-1mm} \prod_{i=1}^{M} \psi_i^{K-M+1} (K\hspace{-1mm}-\hspace{-1mm}i)!}
\end{align}
where \( \Phi(K\tau) \) is an \( M \times M \) matrix with elements
\begin{align*}
    \Phi_{i,j}(x) = \psi_j^{K - i + 1} \, \gamma \left( K - i + 1, x/\psi_j \right)
\end{align*}
and \( \psi_1 > \dots > \psi_M \), with \( \psi_m = w_m (\sigma_s^2 |\alpha_m|^2 + \sigma_n^2) \).
\section{Numerical Results}

This section presents the results of numerical simulations conducted with an UPA antenna with isotropic elements, configured with five \textsf{PU}s (\(Q=5\)) and multiple \textsf{SU}s. The UPA operates with a transmission power of \( p = 26.98\,\text{dBm} \), an antenna gain of \(G =5\,\text{dB} \), and a noise power of \( -60\,\text{dBm} \) \cite{Zhang2024icc,zhang2024optimal}. The antenna is positioned \( 60\,\text{m} \) above ground level, and the transmission frequency is set to \( 5.2\,\text{GHz} \). 
For simplicity, we assume a perfect reporting channel between each \( \textsf{SU}_m \) and the FC, allowing the focus to remain on the study’s core objectives. Detection is performed using \( K = 100 \) samples collected within the channel coherence time of a Rician channel. The simulations consider both static and mobility scenarios for the \textsf{SU}s. Results are presented as receiver operating characteristic (ROC) curves. 

Under the static scenario, the \textsf{SU}s are positioned at fixed locations. Fig.~\ref{ROC_data_fusion} shows the ROCs for WED and WEVD with $M=10$ and $L=64$. Analytical results, derived using \eqref{pfwed}, \eqref{pdwed} for WED and \eqref{pfevd}, \eqref{pdevd} for WEVD, show strong agreement with simulations, validating the analytical accuracy. Notably, WED and WEVD exhibit similar performance. The plot also includes results for uniformly weighted ED~\cite{atapattu2011energy} and EVD~\cite{chiani2017probability}, where the proposed WCSS demonstrates superior performance.
For example, at \( P_f(\tau) = 0.1 \), the \( P_d(\tau) \) of WED and WEVD is approximately twice that of their uniformly weighted counterparts.
Next, Fig.~\ref{ROCdata_users} compares the detection performance for different \textsf{SU} counts (\( M = 5, 10 \)) placed at various locations along roads. As expected, increasing \( M \) enhances detection performance due to the higher likelihood of more \textsf{SU}s being positioned in areas with stronger signal strength. Similar to the previous case, both WCSS detectors, WED and WEVD, exhibit comparable performance.  

Under mobility scenario, \textsf{SU}s travel along roads at \( 10\,\text{m/s} \), starting from the same initial positions as in the static case. The weights of each \textsf{SU} are updated every $\mu$-second interval as $r_m$ and $AF(\theta_m, \phi_m)$ change over time. Fig.~\ref{ROCdata_mobility} shows that the ROC performance declines as $\mu$ increases. This occurs because larger $\mu$ values cause \textsf{SU}s to use outdated weights, introducing errors that degrade the ROC performance. 

Finally, Fig.~\ref{ROCdatadecisionvant} compares the detection performance for antenna configurations with \( L = 64 \) and \( L = 128 \), using \( M = 10 \). Interestingly, reducing \( L \) improves performance, contrary to traditional MIMO systems where location-based directive beamforming is not feasible. Specifically, a smaller \( L \) broadens the main and side lobes, increasing coverage and enhancing the received signal strength for most \textsf{SU}s, as shown in Fig.~\ref{Heatmap16}. In contrast, a larger \( L \) sharpens the lobes, focusing more narrowly on the intended \textsf{PU}, as illustrated in Fig.~\ref{Heatmap128}. However, this behavior is context-dependent, as detection performance varies with the spatial distribution of \textsf{SU}s relative to the \textsf{PU}. In certain cases, a larger \( L \) may outperform a smaller \( L \) when more \textsf{SU}s are positioned in regions with higher received signal strength. 

\section{Conclusion}
This paper introduces a novel UPA-based weighted cooperative spectrum sensing (WCSS) framework, significantly enhancing detection reliability in cognitive radio networks. By leveraging the spatial variation of signal strength across users, we propose a weighting factor for distributed secondary users that incorporates all user-specific parameters. This factor is then applied for commonly used energy detection and eigenvalue detection.
Our analytical results provide novel closed-form expressions for false alarm and detection probabilities for both weighted energy detection and weighted eigenvalue detection using probability and random matrix theory (RMT). Numerical simulations confirm the effectiveness of WCSS, particularly in scenarios with static and dynamic user behavior, highlighting its performance improvements over traditional uniformly weighted detection methods. Future work will optimize weighting factors based on UPA characteristics and detector properties, and extend to dynamic cognitive radio networks, including cellular, vehicular, and satellite systems.
\section*{Acknowledgment}
This work has been supported in part by the SmartSat CRC, whose activities are funded by the Australian Government’s CRC Program; and in part by the Australian Research Council (ARC) Future Fellowship under Grant FT210100728.

\end{document}